# High-Throughput Identification and Statistical Analysis of Atomically Thin Semiconductors

*Juri G. Crimmann,[†] Moritz N. Junker,[†] Yannik M. Glauser,[†] Nolan Lassaline,[†,§] Gabriel Nagamine,[†] and David J. Norris[*,†]*

[†]Optical Materials Engineering Laboratory, Department of Mechanical and Process Engineering, ETH Zurich, 8092 Zurich, Switzerland

[§]Department of Physics, Technical University of Denmark, 2800 Kongens Lyngby, Denmark

ABSTRACT. Transition metal dichalcogenides (TMDs) are layered two-dimensional semiconductors explored for various optoelectronic applications, ranging from light-emitting diodes to single-photon emitters. To interact strongly with light, such devices require monolayer TMDs, which exhibit a direct bandgap. These atomically thin sheets are typically obtained through mechanical exfoliation followed by manual identification with a brightfield optical microscope. While this traditional procedure provides high-quality crystals, the identification step is time-intensive, low-throughput, and prone to human error, creating a significant bottleneck for TMD research. Here, we report a simple and fully automated approach for high-throughput identification of TMD monolayers using photoluminescence microscopy. Compared to a manual search and verification, our methodology offers a four-orders-of-magnitude decrease in the time a researcher must invest per identified monolayer. This ability enables us to measure geometric and photoluminescence-intensity features of more than 2,400 monolayers and bilayers of $WSe_2$, $MoSe_2$, and $MoS_2$. Due to these large numbers, we can study and quantify material properties previously inaccessible. For example, we show that the mean photoluminescence intensity from a monolayer correlates with its size due to reduced emission from its edges. Further, we observe large variations in brightness (up to 10×) from $WSe_2$ monolayers of different batches produced by the same supplier. Therefore, our automated approach not only increases fabrication efficiency but also enhances sample quality for optoelectronic devices of atomically thin semiconductors.

# INTRODUCTION

Materials science is becoming increasingly automated with the introduction of robotic systems,[1,2] machine learning,[3-7] and artificial intelligence.[8] Scientists are leveraging these advances to spend more time on cognitively critical tasks to increase research efficiency. Fully automated research[9] is already employed in synthesis-based[10-12] and fabrication-focused fields.[2,13] For example, new semiconductor[10] and metal[11,12] nanoparticles are synthesized, thin-film processing conditions are optimized,[13] and two-dimensional (2D) heterostructures are assembled.[2]

The latter involves 2D materials such as graphene[14] and transition metal dichalcogenides (TMDs).[15] Like graphene, TMDs are layered materials with covalent bonds within each layer and van der Waals interactions between layers.[16,17] This structure allows the optical and electronic properties of atomically thin semiconductors to be explored. In particular, while typical TMDs are semiconductors with indirect bandgaps in multilayer form, their monolayers have direct bandgaps.[18] Therefore, unlike multilayer TMDs, monolayers exhibit bright photoluminescence (PL).[19] Consequently, TMD monolayers are being investigated for optoelectronic devices, such as light-emitting diodes,[20] photodetectors,[21] and solar cells.[22] Other more subtle effects in TMDs can also be exploited. For example, left- and right-handed circularly polarized light couples to optical transitions in different regions (valleys) of the electronic band diagram. This gives rise to applications in valleytronics, where the valley, as a new degree of freedom for the electrons, is used to carry and store information.[23]

For device fabrication, the material must be obtained with the correct thickness. TMD monolayers can be acquired using several methods.[18] The most prominent strategies are direct growth through chemical vapor deposition (CVD)[24] and mechanical exfoliation.[25] The latter approach was used to obtain graphene from bulk crystals of graphite.[14] Twenty years later, this preparation method remains prevalent for graphene and TMDs because the quality of exfoliated monolayers remains superior to CVD-grown materials.[26]

The mechanical-exfoliation process (Figure 1a,b) involves thinning down a layered bulk crystal with adhesive tape. By bringing a piece of tape into contact with the crystal, a stack of many layers is removed. The thickness is further reduced by transferring the material to another piece of tape, which is repeated several times. Ultimately, monolayers are obtained by pressing the final tape onto a substrate. To confirm successful exfoliation, manual inspection with an optical brightfield (BF) microscope is typically used,[26] exploiting differences in contrast between monolayers and multilayers.

Depending on the size and shape requirements of the desired monolayers, this entire preparation process requires significant manual effort. First, the number of suitable monolayers transferred per unit area of the substrate is low.[26] Second, classification of the flakes with BF microscopy is time intensive due to the poor contrast between multilayer and monolayer TMDs.[27] Finally, flake exfoliation and identification are prone to human error and lead to fatigue for the experimentalist.

To address the critical challenges involved in flake identification, efforts to automate this step have been pursued. This has included classic computer vision[2,28] and more advanced optical-detection algorithms, including those based on neural networks,[8,29-31] machine learning,[3-7,32] and deep learning.[8,29,30,33] Despite these efforts, the majority of TMD work is still done manually.[34] More advanced approaches have not been widely adopted due to their complexity and high implementation costs. Consequently, manual identification of flakes continues to occupy a large fraction of the overall time required for TMD-device production.

Here, we address this problem by providing an accessible approach to identify and classify TMD flakes. Instead of BF microscopy, we utilize PL to easily distinguish monolayers from bilayers. PL offers a significantly improved monolayer-to-background contrast compared to BF. We demonstrate our simple and cost-effective setup by classifying three types of TMDs: $WSe_2$, $MoSe_2$, and $MoS_2$. After collecting more than $5 \times 10^4$ PL images, an automated analysis procedure extracts the geometric and intensity features of 1734 $WSe_2$, 517 $MoSe_2$, and 230 $MoS_2$ monolayers and bilayers. Such large numbers allow us to detect and quantify statistical variations within one type of TMD or between different TMDs. For example, we observe a strong size and PL-intensity correlation for monolayers of all three materials. We attribute this effect to reduced emission from edges. Lastly, we compare monolayers from different growth batches of $WSe_2$, demonstrating a route to preselect flakes based on geometric and PL properties. Thus, our approach enables more efficient monolayer preparation and provides statistical analysis to improve sample quality, leading to faster and better device fabrication.

**RESULTS AND DISCUSSION**

**Brightfield *versus* Photoluminescence.** For this study, we chose $WSe_2$, $MoSe_2$, and $MoS_2$, which are widely explored TMDs for a variety of applications.[20-22,35] All exhibit a trigonal prismatic (2H) phase with hexagonal appearance in plane (see Figure 1a).[18] To prepare monolayers, we performed mechanical exfoliation (Figure 1b) with Scotch tape from a commercially available bulk crystal onto a polydimethylsiloxane (PDMS) substrate supported by a glass slide[36,37] (see Methods and the Supporting Information). We explicitly chose materials, substrates, and preparation methods that are commonly used.

After exfoliation, a big challenge is monolayer identification. In BF, a white-light source illuminates a large area of the substrate, and the reflection is collected (Figure 1c). In this case, the contrast between atomically thin layers and the background is poor. In addition, the BF images are polluted by signals from the substrate, tape residue, dust, and thicker flakes. For example, Figure 1d shows a region from an exfoliated $WSe_2$ sample on PDMS in BF. The initial color image from the camera was converted to grayscale and normalized to values between 0 (minimum, Min) and 255 (maximum, Max). The BF image shows $WSe_2$ flakes of different thicknesses, small bubbles entrapped between the PDMS and the glass slide, and shadows due to reduced reflectivity from the PDMS. Low contrast makes it challenging to identify monolayers. Consequently, researchers often search near a thicker crystal. We used this approach to find a monolayer, which is marked with a dashed line in Figure 1d. Without such a visual aid, the monolayer is barely visible at this optical magnification (10×).

A histogram of intensity values shows the challenge of identifying monolayers. In Figure 1e, a pixel–intensity histogram of the BF image from Figure 1d is shown. The 1280 × 1024 pixels are binned onto an intensity axis, spanning from 0 (Min) to 255 (Max). Pixels that contain the monolayer are plotted in light blue while all other pixels are shown in dark blue (which dominates the histogram). Many pixels have the same intensity value as those of the monolayer, obscuring its signal. Most monolayer pixels have intensity values near a specific bin (light blue arrow, Figure 1e). The inset shows just the top part of this bin, where some signal from the monolayer can be seen (light blue), clearly highlighting the difficulty of using BF.

To circumvent this problem, a common practice is to increase the optical magnification in BF. While this aids in the visual examination of monolayers that have lateral dimensions on the micrometer scale, higher optical magnification also reduces the field of view. This tradeoff severely limits the overall throughput of the procedure.

An alternative solution is to exploit the bright PL of TMD monolayers,[15,19] which can potentially provide a more efficient identification strategy.[28,38] To explore this possibility, we excite the sample with the same white-light source used in BF. However, additional optical elements must be added to distinguish the PL signal from the broadband illumination. Therefore, we introduce shortpass and longpass filters in the excitation and collection paths, respectively. This configuration then allows us to detect the PL (Figure 1f). Figure 1g shows a PL image for the same monolayer shown in Figure 1d. As before, it is rescaled to values between 0 (Min) and 255 (Max). Unlike in BF, the PL image shows a uniform and relatively dark background. Hence, the monolayer can be clearly seen, while thicker crystals, bubbles, *etc.*, are not visible.

This is quantified in Figure 1h, which shows the histogram of the PL image (Figure 1g), plotted as in Figure 1e. The histogram clearly reveals the advantage of PL for the identification of monolayers compared to BF. The only bright pixels in the PL image are from the monolayers, which are obscured in the BF image (Figure 1e). Moreover, a threshold value can be set for easy detection of monolayers.

**Automated Identification of TMD Monolayers with PL.** This process can easily be automated. Figure 2a shows a simplified sketch of the optical setup. (A more detailed schematic and a parts list are provided in the Supporting Information.) The emission from a broadband light-emitting diode (LED) is sent through a shortpass filter and coupled into the microscope with a beamsplitter to illuminate the sample *via* a 10× objective. The reflection from the sample, which is placed on a motorized stage, is collected with the objective and passes the same beamsplitter. In a typical BF microscope, the reflected light would be directed to a color camera. While maintaining this option, we introduce a second beamsplitter with a longpass filter to collect PL images on a second gray-scale camera. A similar setup has been used to inspect PL from CVD-grown TMDs.[28,38] However, our configuration enables the simultaneous detection of BF and PL information, which we apply to high-quality mechanically exfoliated TMDs.

The working principle of our automation is depicted in Figure 2b. It involves cycling through four primary steps. Step 1 is the acquisition and processing of an image. A WSe$_2$ monolayer on PMDS is shown as an example. In step 2, the intensity distribution is analyzed. An automated threshold value separates the background and monolayer pixels in the histogram (plotted as in Figure 1h). The number of pixels above the threshold is then calculated. The system decides that an image contains a monolayer when the number exceeds a user-defined value, which also corresponds to an area (which below we refer to as the "size"). When a monolayer is detected, the image is tagged, and specific data such as the position and size are saved, as illustrated in step 3. The stage is then moved to the next position in step 4, and the entire process is repeated until the desired area is scanned.

**Proof of Concept: Finding TMD Monolayers with PL.** Figure 3a–c shows an example BF image, PL image, and the corresponding PL pixel–intensity histogram for WSe$_2$ on PDMS. The BF image is white-balanced and its intensity was rescaled to fit the entire range. The PL image was adjusted as in Figure 1g. In Figure 3a, one could mistakenly identify the large bilayer at the top of the thicker crystal as a monolayer. However, this region appears dim in the PL image (Figure 3b), while an actual monolayer exhibits significantly brighter PL. Notably, the small monolayer cannot be seen in the BF image at this magnification, as the contrast is too low.

The corresponding histogram of intensity values in Figure 3c shows three different regions, separated by vertical dashed lines. The darkest region (left of the first line) is the background. Between the lines, pixels of the bilayer can be found. The brightest region, which is to the right of the second line, represents pixels of the monolayer. We confirmed this by Raman measurements (Figure 3d). In contrast to the monolayer, the bilayer exhibits a Raman peak at 308 cm$^{-1}$, consistent with previous reports.[19,39]

We collect similar data for MoSe$_2$ (Figure 3e–g) and MoS$_2$ (Figure 3h–j), illustrating the general applicability of our approach. For both materials, the enhanced contrast provided by PL allows the detection of monolayers barely visible in BF. Raman measurements again confirm the detection of monolayers (Figure 3d).[19,39,40] Unlike WSe$_2$, we could not detect bilayers for MoSe$_2$ and MoS$_2$ due to their weak emission. Nevertheless, bilayers are more emissive than thicker crystals, and we expect that they could be detected with a more sensitive camera.[19,41] We observed that the PL intensity of the monolayers decreased from WSe$_2$ to MoSe$_2$ to MoS$_2$, which influenced the acquired PL images. In particular, the background is most apparent in MoS$_2$. There, higher camera gains and exposure times give rise to camera artifacts, and thicker crystals become visible as dark shadows. The differences in monolayer brightness are also reflected in the corresponding histograms (Figure 3c,g,j), where the dashed threshold lines are shifted towards higher intensities for MoSe$_2$ and MoS$_2$, indicating a brighter background compared to the monolayer.

Figure 3k demonstrates the overall performance of our automated detection for the three materials. We prepared 10 TMD-on-PDMS samples for each (see Methods). The automation program ran for 34 min per 12 mm × 12 mm scan for each sample (the scan time). The exchange to the next sample and restart of the program took less than 15 s (the operator time). We found, on average, 40 ± 13 WSe$_2$ mono- and bilayers, 52 ± 23 MoSe$_2$ monolayers, and 23 ± 13 MoS$_2$ monolayers per sample (with the range denoting one standard deviation). These values are shown as bars in Figure 3k. The yield of monolayers is the highest for MoSe$_2$, resulting in the lowest average operator time per monolayer of around 0.3 s.

**High-Throughput Characterization of TMD Monolayers.** Due to the aforementioned challenges in monolayer identification and their integration, TMD studies are typically limited to a few flakes or devices.[35,41] Our approach provides an efficient method of TMD identification, enabling larger datasets. To demonstrate this, we analyzed a total of 20,700 PL images of 10 WSe$_2$, 10 MoSe$_2$, and 10 MoS$_2$ samples. An area of 12 mm × 12 mm from each sample was scanned, giving a total scanned area of $4.3 \times 10^3$ mm$^2$. We identified individual monolayers (as well as bilayers for WSe$_2$) in each PL image with a Python-based script and extracted the size, mean intensity, perimeter, and axes (major and minor) of a fitted ellipse. Below we refer to the identified monolayers and bilayers collectively as flakes.

Figure 4a,b shows size and aspect-ratio distributions, Figure 4c reveals the correlation between these two geometric features, and Table 1 provides a statistical overview for all of the identified flakes. The size distribution of $MoS_2$ is notably shifted to larger sizes compared to $WSe_2$ and $MoSe_2$. The aspect ratios, which were calculated by dividing the major by the minor axes, are comparable for all materials and scale with size. The similarity of the scatter plots in Figure 4c indicates that the exfoliation leads to flakes with comparable morphologies for all materials.

Because both geometric and PL information of the TMD flakes is available, correlations between them can also be analyzed. Figure 4d plots mean intensity *versus* size for all materials. The mean intensity is determined by averaging the PL signal for all pixels within a flake. One might expect that the average value would be independent of size, but $WSe_2$ and $MoSe_2$ show a strong correlation between size and mean intensity. $MoS_2$ has a similar correlation, but it is weaker, which we attribute to the lower PL intensity of $MoS_2$.

Notably, the mean-intensity data for $WSe_2$ can be divided into two regions (shaded in Figure 4d). The data from the upper region shows a strong correlation between mean intensity and size, while that from the lower stays just above the minimum mean-intensity value. We investigated 10 flakes from each region *via* Raman spectroscopy (Figure S7). These measurements reveal that the upper region consists of monolayers, while the lower features bilayers. Because we can only detect bilayers of $WSe_2$, we did not observe this behavior for the other materials.

To understand the strong correlation between size and mean intensity for the monolayers, we can look at a single $MoSe_2$ flake (Figure S8). The PL emission from the center of the flake is significantly brighter than from the edges. Because the relative contribution from the flake edge decreases with increasing flake size, this observation provides a potential explanation for the correlations in Figure 4d. To test this hypothesis, Figure 4e plots the mean intensity as a function of the perimeter-to-size ratio. We observe an anticorrelation for all three materials, which indicates that the PL intensity differences between the center and edges of the flakes cause the observed correlations between size and mean intensity. Such differences were previously observed and attributed to variations in the defect density.[42-46] Our results are then consistent with higher defect densities near the flake edges.

**Preselection and Screening Capabilities.** Above we exploited our method to compare different TMD materials. It revealed intrinsic differences and similarities, such as the correlation between mean PL intensity and size. Alternatively, our method can compare different batches of the same material to find the best-suited flakes for a given application. We investigated five different batches of $WSe_2$ from the same supplier. From these bulk crystals, we prepared 50 samples, ten from each batch, and scanned a total area of $7.2 \times 10^3$ mm$^2$. We extracted geometric and PL data as in Figure 4. In total, 34,500 PL images were analyzed, and 1734 $WSe_2$ flakes were identified.

Figure 5a plots the mean PL intensity *versus* size for all five batches. The intensity was corrected to compare the flakes (see Methods). Importantly, we observed strong variations between batches of the same material. For example, the brightest flake of batch 1 is more than 10× brighter than that of batch 5. Batches 2 through 4 show intensities in between. All batches show a similar mean intensity *versus* size correlation, where large flakes are brighter than small flakes. Figure 5b shows the corrected mean-intensity distributions for all batches. Batch 1 clearly has many more bright flakes. These differences in PL can arise due to different defect densities.[42,43]

A high-quality starting material is of great importance for many applications. For example, efficient emission is beneficial for LEDs.[47] In our study, batch 1 has flakes with superior PL properties and is the candidate of choice. Without this information, an experimentalist must rely purely on luck. To be more deterministic, our automated PL identification method can be employed.

**CONCLUSIONS**

We have presented a simple approach based on PL microscopy for the high-throughput identification of TMD monolayers. This methodology reduces the time an experimentalist must invest per monolayer by four orders of magnitude, which enables the acquisition of larger datasets. Our ability to collect more data allows us to observe statistical correlations, *e.g.*, between the mean PL intensity and the size of a monolayer for $WSe_2$, $MoSe_2$, and $MoS_2$. Our analysis indicates that emission from smaller flakes is reduced due to weak PL at their edges. Lastly, we observe differences up to 10× in PL intensity between WSe$_2$ batches from the same supplier, which illustrates the prescreening capabilities of our method. Such high differences in the PL intensity additionally showcase the necessity of our approach to obtain the best flakes for devices.

Automatically identifying TMD monolayers through their bright PL is simple and accessible. In fact, using the provided automation code, an existing BF setup can easily be adapted by adding a pair of optical filters and a motorized stage to the sample holder. Additionally, the setup could be expanded further to acquire PL spectra, lifetimes, *etc.*, for a thorough prescreening process. Our method could also be applied to other TMDs. For example, monolayers of MoTe$_2$ exhibit PL many times brighter than multilayers.[48] Using the appropriate optical elements and detectors for near-infrared emission, we expect that monolayers of MoTe$_2$ could also be identified. More generally, fast characterization is essential for all exfoliated monolayers. Here, our method addresses this issue, by encouraging the TMD community to use both brightfield and photoluminescence microscopy.

**METHODS**

**Sample Preparation.** Flux-grown WSe$_2$, MoSe$_2$, and MoS$_2$ bulk crystals were purchased from 2D Semiconductors, and PDMS (PF-40X40-065-X4) from Gel-Pak. Samples were prepared through mechanical exfoliation (see Figure S1). 10 samples were prepared simultaneously, which allowed them to be compared. Preparation was done outside the glovebox in ambient conditions. 10 glass slides and 15 pieces of Scotch® Magic™ Tape (each 15 cm long) are placed on paper (Figure S1a). Dust is removed from the glass slides by blowing with dry nitrogen. The tapes are numbered 1–15. 10 patches of PDMS are cut to a size of 15 mm × 15 mm and placed slowly on the glass slides to avoid air pockets (Figure S1b). A bulk TMD crystal is then placed on tape 1 (Figure S1c,d). This tape is folded together so that both sides of the crystal are in contact with its sticky side. It is subsequently unfolded, and the remaining bulk crystal is removed with tweezers. Two areas of transferred material are observed on the tape (Figure S1e). The edges of the PDMS patches are secured to the glass slides with two additional pieces of tape (Figure S1f). From tape 1 onwards, the exfoliation scheme described in Table S2 is used. Specifically, tape 1 is first picked up, brought in contact with tape 2, rubbed lightly with the thumb, and removed slowly and steadily. These steps are repeated according to the scheme until all 15 pieces of tape have been used. If 50% of the material is transferred at each exfoliation step, the scheme leads to 1/16th of the initial material from tape 1 on each of tapes 2 through 15, and 1/8th remains on tape 1. Material coverage was similar for all TMDs used. Tapes are shown in Figure S2 for $WSe_2$, $MoSe_2$, and $MoS_2$. Tapes 6 through 15 are pressed lightly on the prepared PDMS patches. They are rubbed with a finger and a cotton swab (Figure S1g). After 10 min, the tapes are slowly and steadily removed from the PDMS, and some material is transferred (Figure S1h).

**Photoluminescence Microscope.** A detailed schematic of the complete optical setup is provided in Figure S3, and a full parts list in Table S1. In the schematic, the excitation and collection paths are drawn in yellow and orange, respectively. From the latter, the red PL path splits off. A white LED is used as the excitation source. After a 400 nm longpass filter, its light is collimated with lens L1, which is followed by a shortpass filter SP (650 nm for $WSe_2$ and $MoSe_2$; 600 nm for $MoS_2$). A lens L2 reduces excitation losses by narrowing the beam diameter. Mirrors M1 and M2 direct the beam toward lens L3, which is used for Köhler illumination.[49] Next, a 50:50 beamsplitter BS1 is used for $WSe_2$ and $MoSe_2$, while a dichroic mirror DM is used for $MoS_2$. After the light is deflected with mirror M3, it passes a 10× microscope objective MO and illuminates the sample, which is mounted on a stage with motorized actuators MA1 and MA2. These are driven by controllers C1 and C2, which are connected to a computer. Light is collected through MO and then passes M3 and BS1 (or DM) towards a beamsplitter BS2 (or removable mirror RM). BS2 is used when BF and PL are measured simultaneously, while RM is used when the PL signal is low and must be maximized. Lenses L4 and L5 expand the beam for BF detection. The other path, which is used for PL, has a second longpass filter LP2 (for $WSe_2$ and $MoSe_2$) or a bandpass filter BP (for $MoS_2$), followed by mirror M4. As in the BF path, two lenses, L7 and L8, are used to expand the beam. Lenses L6 and L9 focus the light on the color and monochromatic cameras, respectively.

**Microscope Automation.** *Program Architecture.* The architecture of the automation program consists of a combination of threads, classes, signals, and functions in Python (see Figure S4). A multi-threaded architecture was implemented. *Ui_MainWindow* is the main thread that maintains all elements of the graphical user interface (GUI), such as buttons, text boxes, images, and the progress bar. It continuously refreshes the elements of the GUI and responds to clicking events by executing the connected function. For example, a click on *Start Scan* executes the *runGridscan()* function. A new thread, called *GridscanThread*, is started, the *run()* function is executed inside it, and a sequence of functions performs the scanning of a sample. *PyQt5* signals are used to enable the flow of information between threads. For example, when a flake is identified during a scan, the *GridscanThread* notifies the *Ui_MainWindow* with the *refresh_flakelist* signal and orders it to update the flake list. The *Images()* class associated with the *GridscanThread* is responsible for creating image objects. These objects store the image and all information about the identified flake.

*Graphical User Interface.* A GUI (see Figure S5) was implemented to improve user experience and minimize errors. It enhances user interaction with clickable buttons and sliders. Live information, such as detected flakes, is provided to the user during a scan. *Home Stage* is used to find the origin of the stage. *Start Scan* initiates a scan. The *Output Messages* window shows live information. The *Up*, *Right*, *Left*, and *Down* buttons, together with the adjacent slider, allow for manual motion of the stage. The slider defines the step size. The *Flake Image* window shows a clickable list of identified flakes. When a flake number is selected, two images are shown. On the left, a PL image is displayed. On the right, a PL image in which the edges are enhanced is shown. The latter supports the user in distinguishing flakes from other unwanted luminescent materials. *Move Stage to Flake 10x* moves the stage to the currently displayed flake. If the 10× objective is exchanged for a 50× objective, *Move Stage to Flake 50x* moves the stage by a predefined amount from the 10× position to correct for an offset between the two objectives.

*Scanning Pattern.* Our stage has maximum lateral dimensions of 12 mm × 12 mm. The scanning area is divided into a grid of rectangles. Each measures 416 $\mu$m × 520 $\mu$m and slightly overlaps with its neighbors. Figure S6 shows a cartoon of the pattern and a photo of the stage.

*Code Availability.* The full Python code for the automated flake detection is available at https://github.com/JGCrimmann/TMDFlakeFinder.

**Flake Analysis.** PL images are processed with the *scikit-image* library in Python. First, for a given PL image, the median PL intensity [med($I$)] and the standard deviation of the background noise ($\sigma_b$) are computed. These values are then used to binarize the same image by applying the following threshold: $\text{med}(I) + 6\sigma_b$. With the functions *skimage.measure.label* and *skimage.measure.regionprops*, the connected white pixels in the binarized image are identified as a flake. To better define its edges, a second threshold value, given by $[\text{med}(I) + \max(I_{\text{flake}})]/2$ is applied, where $\max(I_{\text{flake}})$ is the maximum intensity from the flake. After the second thresholding, each flake is analyzed again with *skimage.measure.label* and *skimage.measure.regionprops*. The extracted properties (size, position, *etc.*) are stored in a data frame and used for further analysis. As a final step, all data is manually validated. An example of the thresholding is shown in Figure S9.

**Comparison of Five WSe$_2$ Batches.** PL data was acquired with exposure times of 0.2 s and 0.8 s for batch 1 and batches 2–5, respectively. The mean PL intensity was rescaled for the different exposure times using [(mean intensity − dark counts) / correction factor]. The correction factor was 1 for batch 1 and 4 for batches 2–5. Afterwards, the corrected mean intensity data was normalized.

**Raman Measurements.** The flakes were characterized using an inVia Renishaw Raman microscope. All measurements were done with a 50× objective (numerical aperture of 0.75) and a 2400 lines/mm grating. For WSe$_2$ and MoSe$_2$, we used less than 0.35 mW from a 514 nm excitation laser, and the sample was exposed for 60 s. We inspected MoS$_2$ with 488 nm laser light with powers below 0.16 mW for 300 s.

ASSOCIATED CONTENT

**Supporting Information**

Additional figures and tables describing the sample preparation, the microscope, the automation, the flake analysis as well as extended Raman and PL measurements.

AUTHOR INFORMATION

**Corresponding Author**

*Email: dnorris@ethz.ch

**ORCID**

Juri G. Crimmann: 0000-0002-0367-5172

Moritz N. Junker: 0009-0006-7769-636X

Yannik M. Glauser: 0000-0002-5362-0102

Nolan Lassaline: 0000-0002-5854-3900

Gabriel Nagamine: 0000-0002-4830-7357

David J. Norris: 0000-0002-3765-0678

**Note**

The authors declare no competing financial interest.

ACKNOWLEDGMENTS

This project was funded by ETH Zurich. N.L. gratefully acknowledges funding from the Swiss National Science Foundation (*Postdoc Mobility* P500PT_211105) and the Villum Foundation (*Villum Experiment* 50355). We thank J. T. Held, A. C. Hernandez Oendra, J. J. E. Maris, and D. Thureja for stimulating discussions and A. Cocina, C. R. Lightner, and S. A. Meyer for technical assistance.

# Tables

|  | WSe$_2$ | MoSe$_2$ | MoS$_2$ |
|---|---|---|---|
| Number of Flakes | 397 | 517 | 230 |
| Largest Flake ($10^4$ $\mu$m$^2$) | 2.8 | 0.3 | 0.8 |
| Average Size ($10^2$ $\mu$m$^2$) | 1.9 | 0.9 | 3.2 |
| Highest Aspect Ratio | 12.6 | 16.3 | 9.0 |
| Average Aspect Ratio | 2.6 | 2.6 | 2.6 |

**Table 1.** An overview of statistical quantities related to the data presented in Figure 4.

# Figures

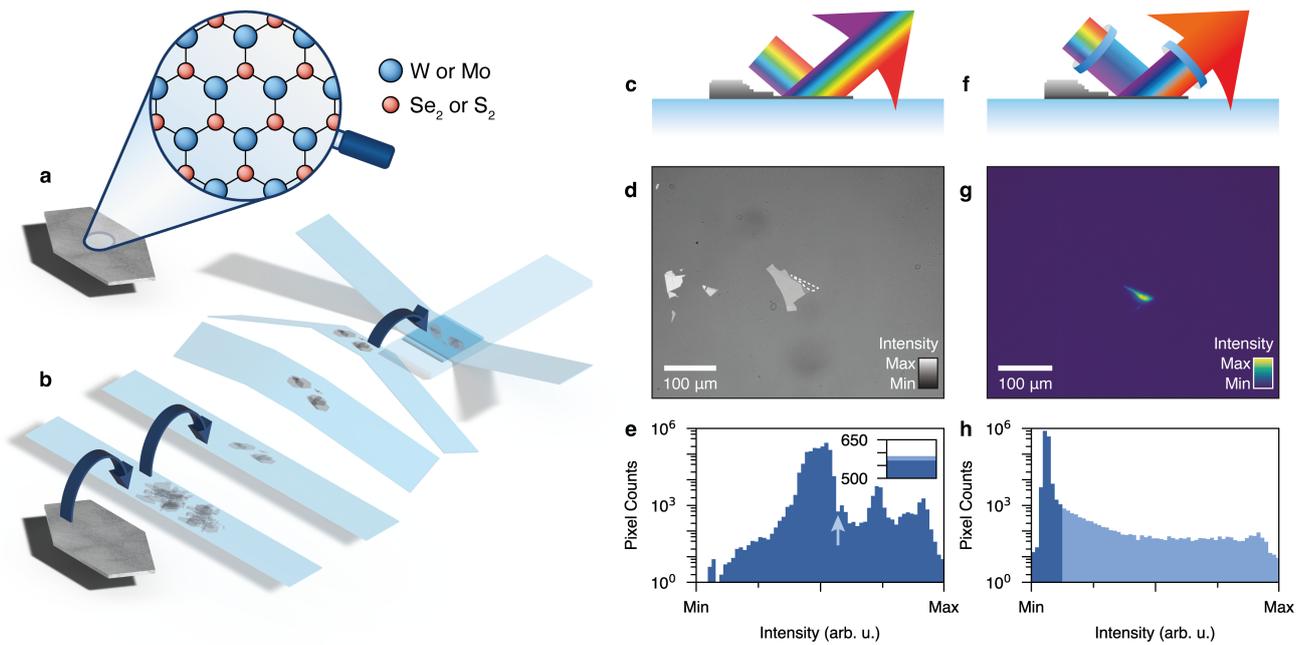

**Figure 1.** Exfoliation procedure and comparison of brightfield (BF) and photoluminescence (PL) microscopy. (a) Cartoon of a TMD bulk crystal depicting the top view of the crystal structure. The legend indicates the atomic constituents. $WSe_2$, $MoSe_2$, and $MoS_2$ are studied in this work. (b) Cartoon showing the process of mechanical exfoliation. Material is thinned down using multiple adhesive tapes and transferred onto PDMS on a glass slide. (c) Cartoon of BF microscopy. Light of the entire visible spectrum strikes a TMD monolayer on PDMS and is collected in reflection. (d) BF microscopy image of $WSe_2$ crystals on PDMS. A monolayer attached to a thicker crystal is highlighted with a dashed white line. (e) Histogram of the number of pixels with a given intensity value in the BF image in (d). The dark blue bins are due to background pixels; the monolayer data is shown in light blue (barely visible). As discussed in the main text, the monolayer contributes to the bins near the light blue vertical arrow. The monolayer contribution at this arrow is highlighted in the inset. (f) Cartoon of PL microscopy. Short-wavelength light excites a TMD monolayer on PDMS. The PL emission is collected by using a long-pass filter. (g) PL microscopy image of the same monolayer as in (d). (h) Histogram of the PL image from (g), plotted as in (e).

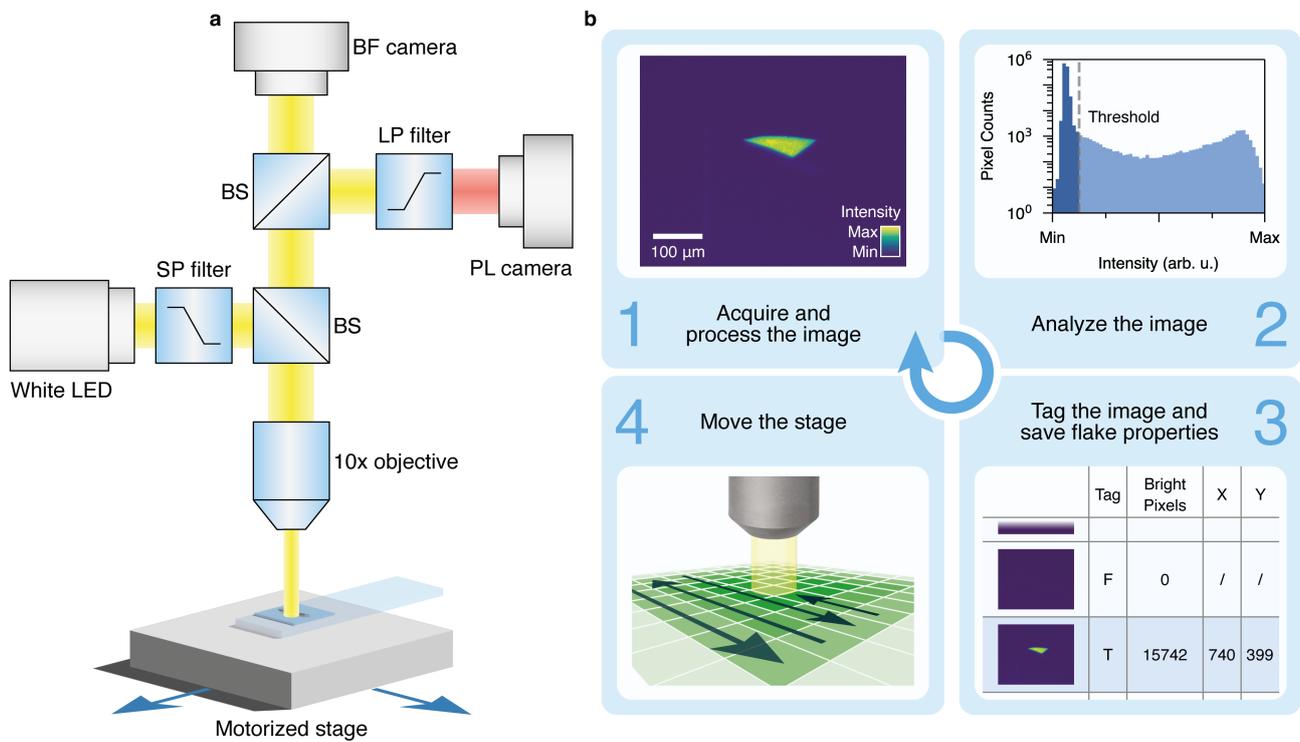

**Figure 2.** Automated identification of TMD monolayers with PL. (a) Schematic of the PL microscope. The output from a white LED is sent through a shortpass (SP) filter and coupled to a 10× objective *via* a beam splitter (BS) to excite a sample mounted on a motorized stage. The reflection is collected through the objective and passes the first BS. A second BS allows both brightfield (BF) and PL information to be collected on two separate cameras. A longpass (LP) filter is required in the second path to eliminate excitation light. See Methods for further details. (b) Working principle of the automation. First, a PL image is acquired and processed. The inset shows a PL image of a $WSe_2$ monolayer on PDMS. Second, the PL image is analyzed. The number of pixels with intensities exceeding a threshold is calculated. The inset shows a histogram of the number of pixels with a given intensity value in the PL image from the first step. The dark blue bins are due to background pixels; the monolayer data is shown in light blue. The intensity threshold is drawn as a grey dashed vertical line. Third, if a flake is detected, the image is tagged, and flake properties are recorded. Fourth, the stage is moved to the next position. These four steps are then cycled.

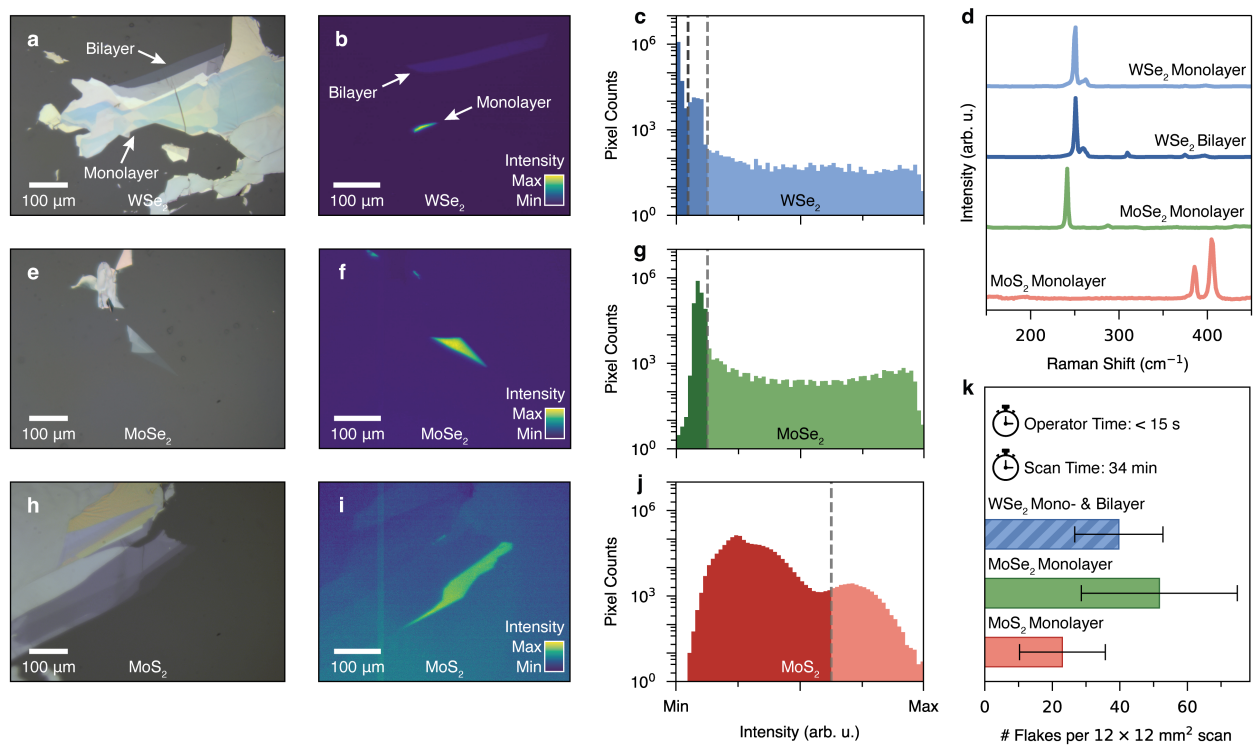

**Figure 3.** Overview of data from investigated materials. (a) BF image of exfoliated WSe$_2$ on PDMS. Locations of a bilayer and monolayer (not visible) are marked with arrows. (b) PL image of the same area shown in (a). (c) Histogram of the number of pixels with a given intensity value for the PL image shown in (b). (d) Raman measurements of the flakes shown in (b, f, i). (e–g) Data for MoSe$_2$ as in (a–c). (h–j) Data for MoS$_2$ as in (a–c). The histograms are plotted as in Figure 2b, with the background data depicted in dark blue, dark green, and dark red, and the monolayer data in light blue, light green, and light red for panels (c), (g), and (j), respectively. Bilayer intensities are also highlighted in medium blue in (c). (k) Overview of the process throughput, obtained by averaging scans from 10 samples for each material. An area of 12 mm × 12 mm per sample was imaged. The average operator time per sample was less than 15 s. The total time per sample was 34 min. The bars show the average number of identified flakes per 12 mm × 12 mm scan, and the error bars indicate the standard deviation.

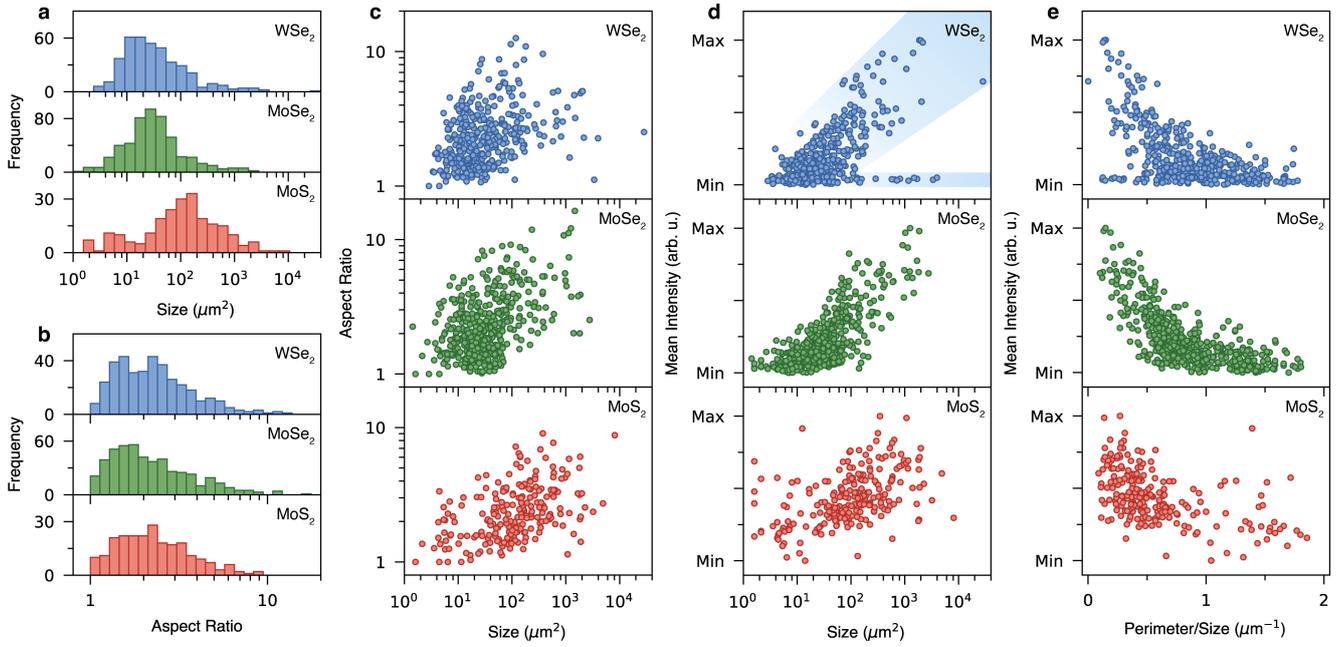

**Figure 4.** Extracted geometric and PL information from exfoliated $WSe_2$ (blue), $MoSe_2$ (green), and $MoS_2$ (red) flakes. 10 samples of each material were prepared. For every sample, an area of 12 mm × 12 mm was scanned. The area of a flake is referred to as its "size." (a, b) Size and aspect-ratio distributions. (c–d) Scatter plots of aspect ratio *versus* size, mean PL intensity *versus* size, and mean PL intensity *versus* the perimeter-to-size ratio, respectively. Each dot corresponds to a single flake. In (d), two regions are shaded for $WSe_2$, as discussed in the text.

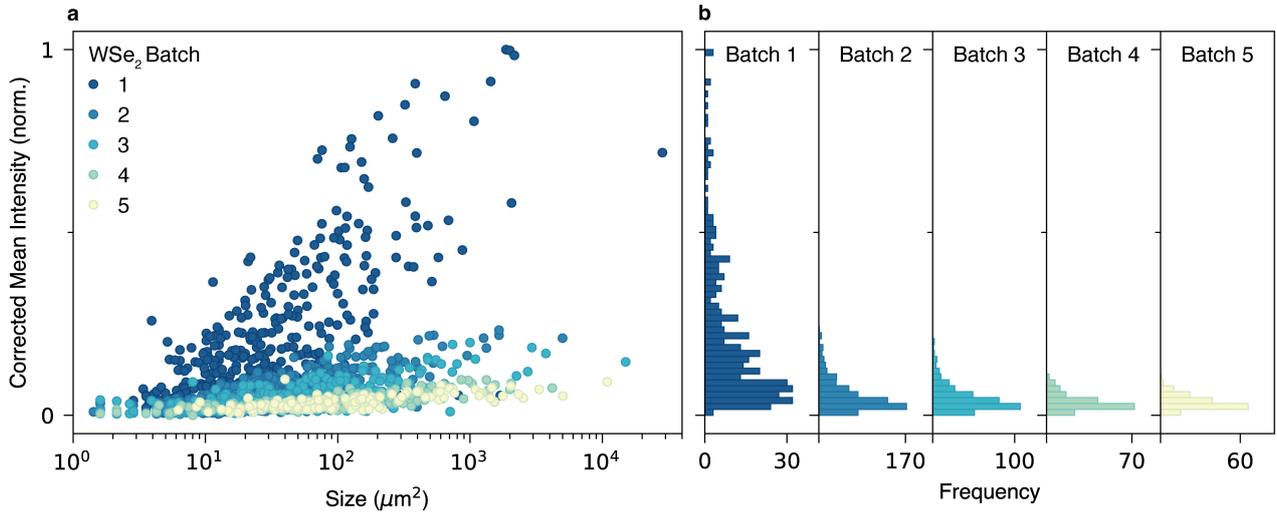

**Figure 5.** Comparison of five $WSe_2$ batches from the same supplier. 10 samples for each batch were prepared by exfoliation. For each sample, 12 mm × 12 mm was scanned. The area of a flake is referred to as its "size." (a) Scatter plots of the mean PL intensity *versus* size for all batches. The intensity was corrected to compare flakes (see Methods). Each dot corresponds to a single flake. (b) Corrected mean-intensity distributions for all batches.

**Table of Contents Graphic**

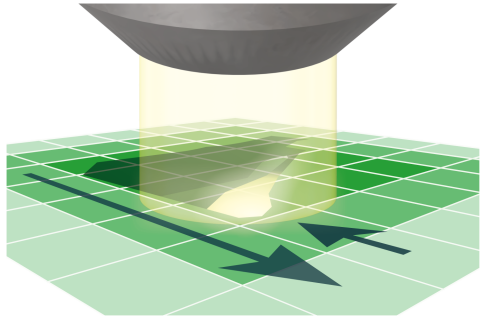

# High-Throughput Identification and Statistical Analysis of Atomically Thin Semiconductors

*Juri G. Crimmann,[†] Moritz N. Junker,[†] Yannik M. Glauser,[†] Nolan Lassaline,[†,§]*
*Gabriel Nagamine,[†] and David J. Norris[\*,†]*

[†]Optical Materials Engineering Laboratory, Department of Mechanical and Process Engineering, ETH Zurich, 8092 Zurich, Switzerland

[§]Department of Physics, Technical University of Denmark, 2800 Kongens Lyngby, Denmark

**Corresponding Author**

\*Email: dnorris@ethz.ch

# S1. SUPPLEMENTARY FIGURES

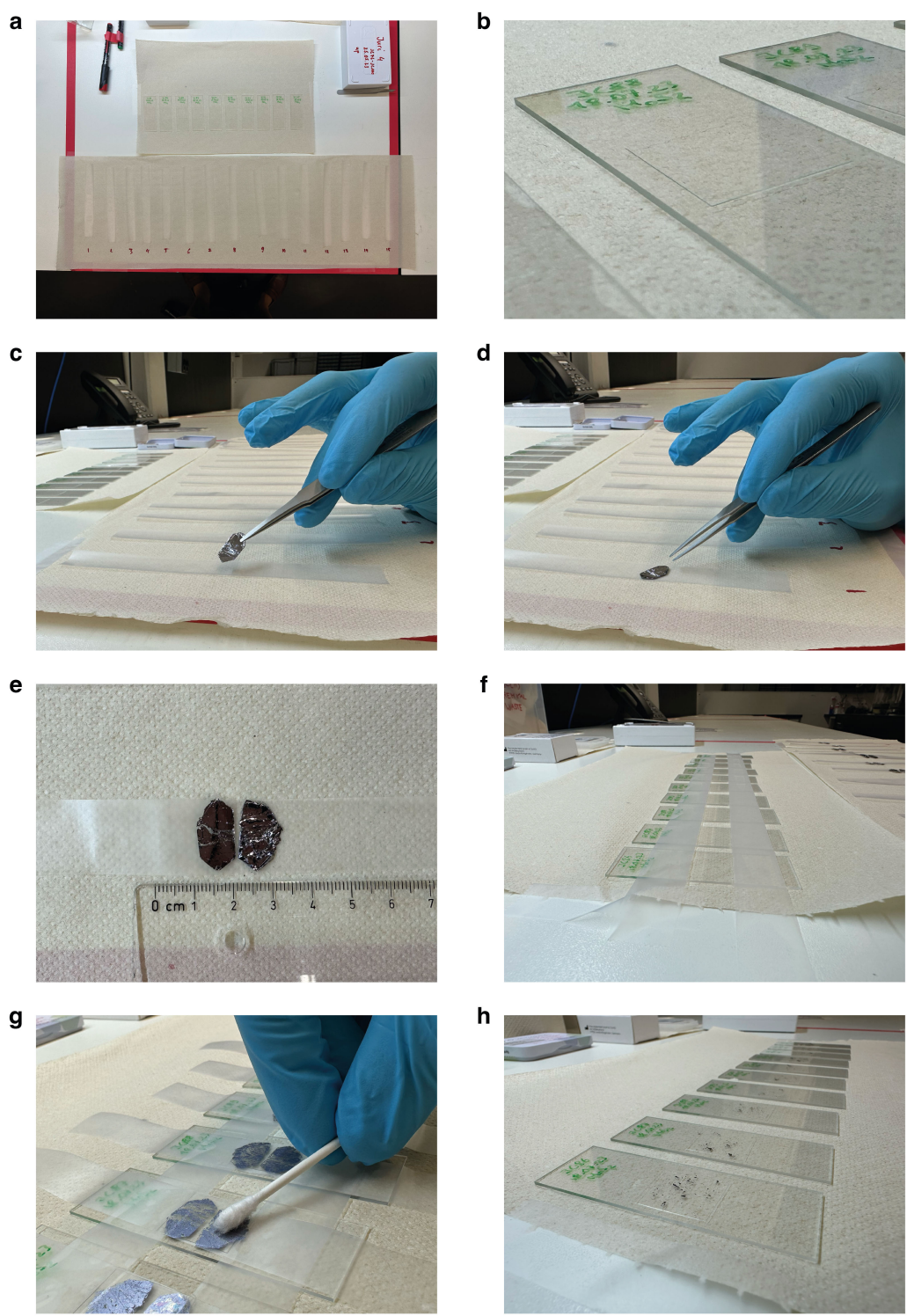

**Figure S1.** Overview of sample preparation. (a) Adhesive tapes and glass slides. (b) PDMS patch placed on a glass slide. (c) Bulk crystal held by tweezers and (d) placed on the tape. (e) Material transferred on tape. (f) PDMS patches secured by tape. (g) In contact with PDMS, tapes are gently rubbed with a cotton swab. (h) Overview after removing the tapes.

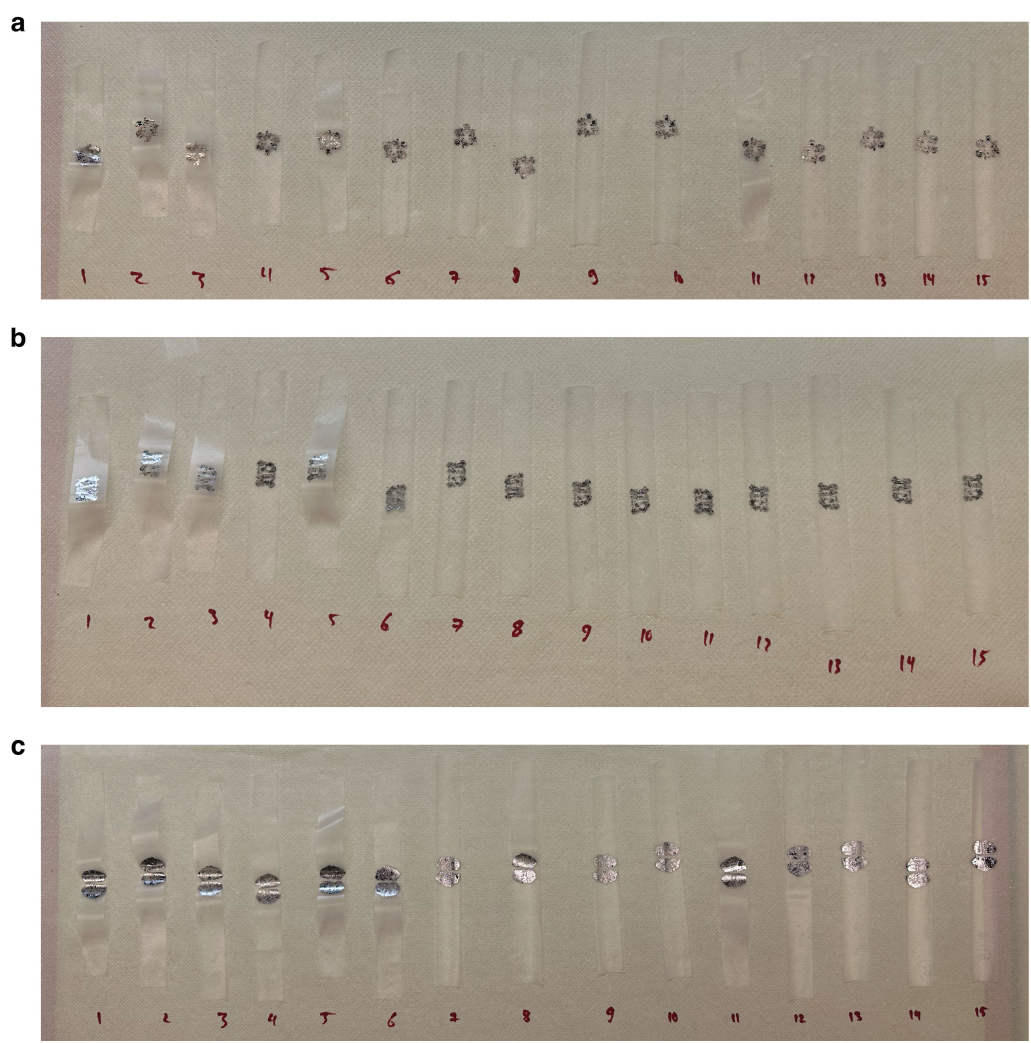

**Figure S2.** Overview of the amount of material used for exfoliation. The photos show tapes prepared according to the scheme of Table S2. The materials are (a) WSe$_2$, (b) MoSe$_2$, and (c) MoS$_2$.

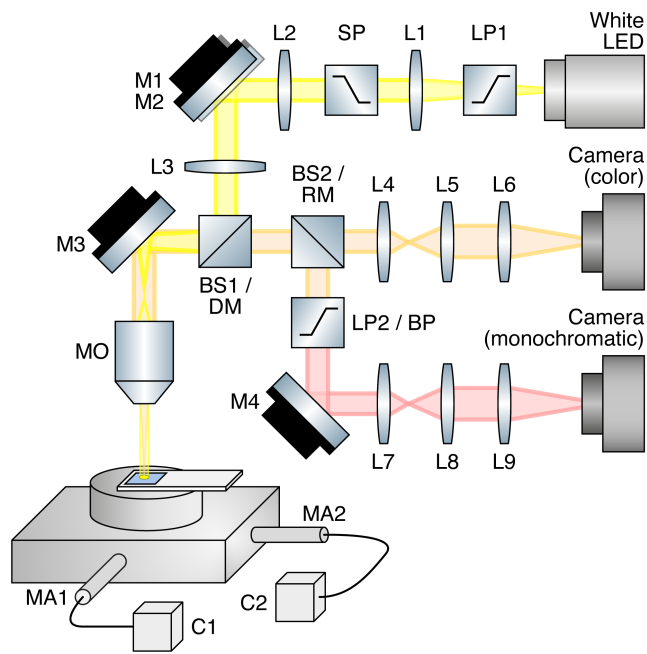

**Figure S3.** Schematic of the complete setup. More information is provided in Methods. All parts are listed in Table S1.

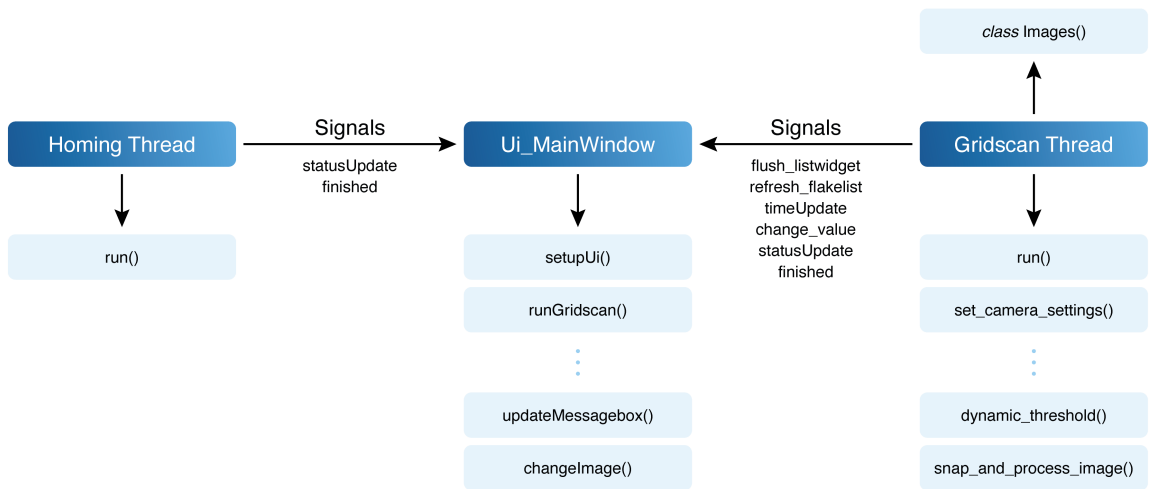

**Figure S4.** The architecture of the automation program. More information is provided in Methods.

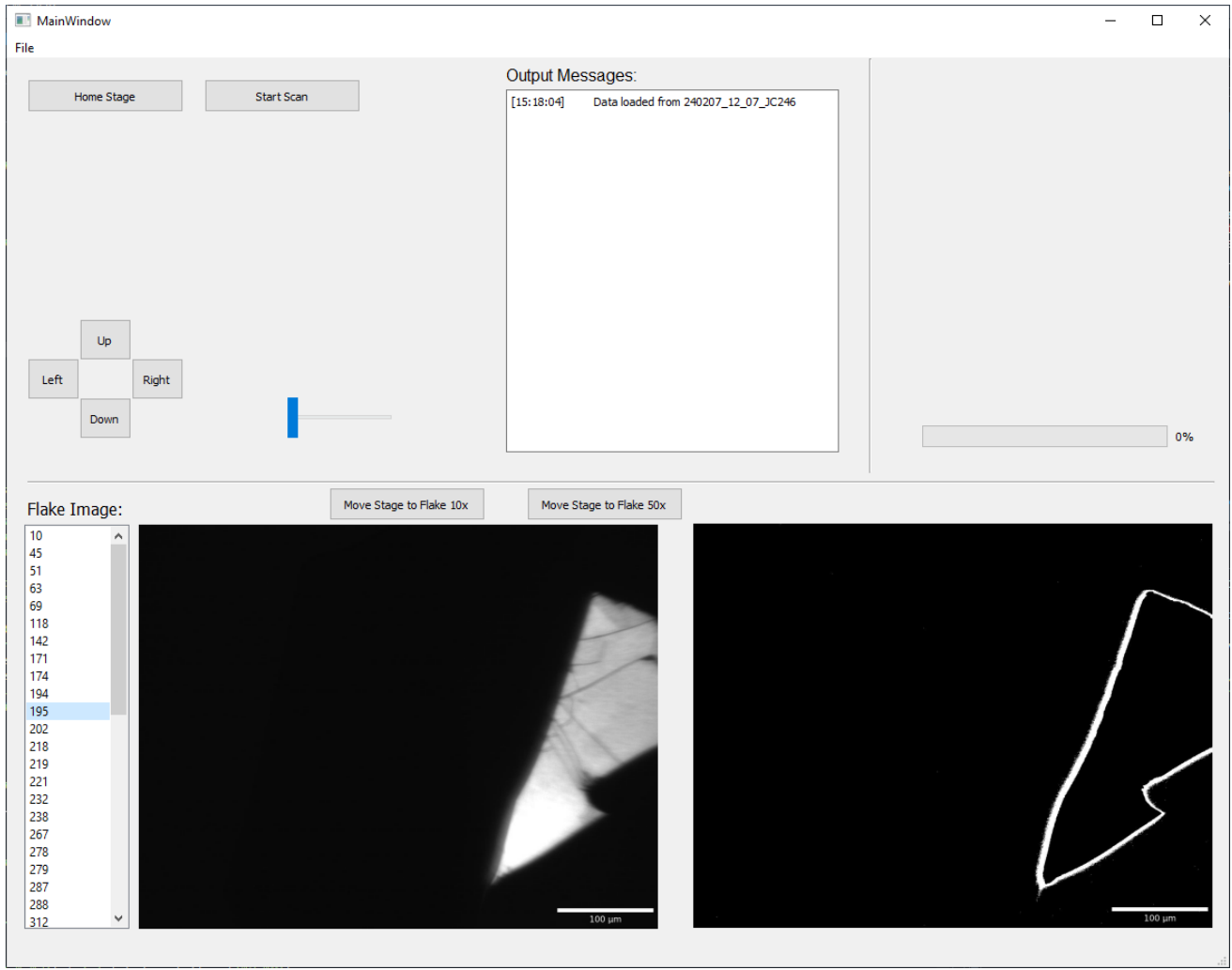

**Figure S5.** Screenshot of the graphical user interface (GUI) of the automation program. More information is provided in Methods.

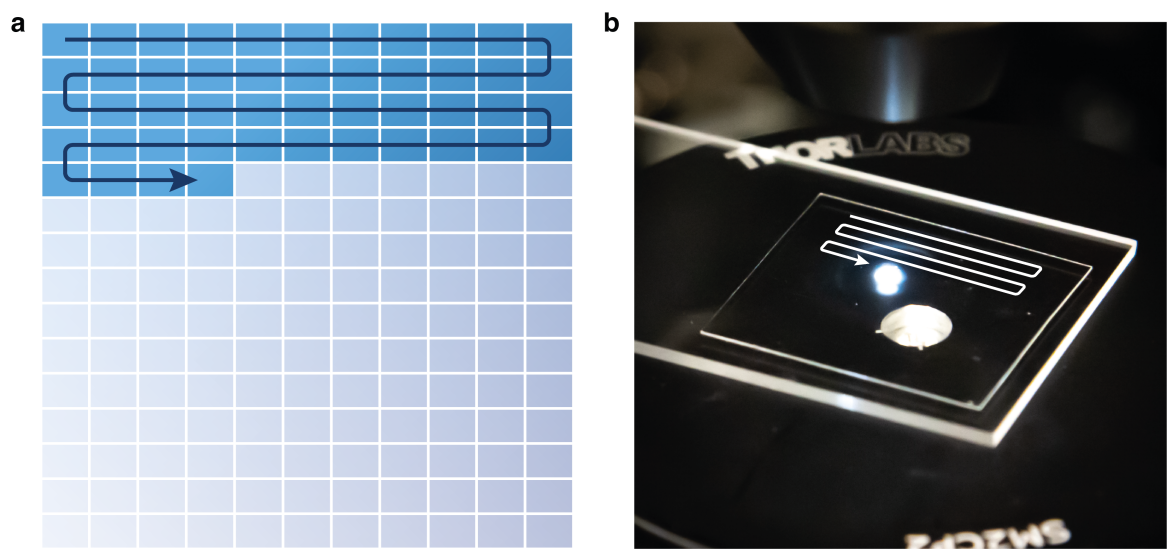

**Figure S6.** Scanning pattern of the automation routine. (a) Cartoon of the scanning pattern. (b) Photo of the stage with an arrow indicating the movement along the patch of PDMS on a glass slide. The bright white dot is the white-light illumination. See also Methods.

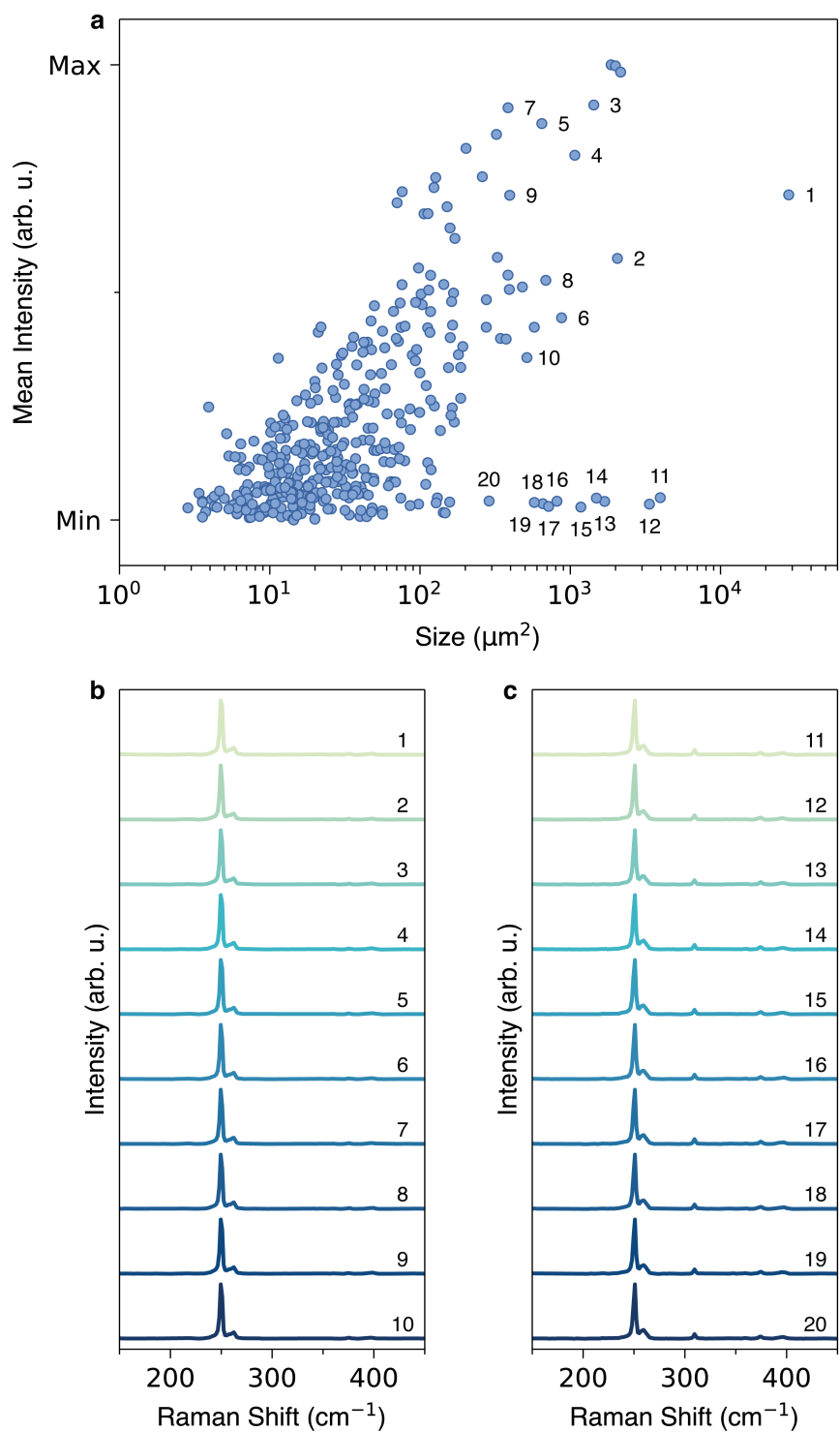

**Figure S7.** Raman spectroscopy analysis. The area of a flake is referred to as its "size." (a) Mean intensity *versus* size correlation of mechanically exfoliated $WSe_2$ mono- and bilayers. Raman spectroscopy data was taken from the numbered flakes and is shown in (b) and (c). See also Methods.

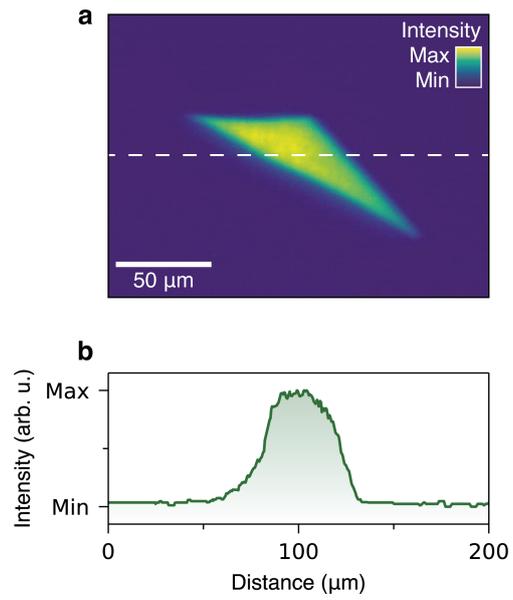

**Figure S8.** Photoluminescence of a MoSe$_2$ monolayer. (a) Photoluminescence image of a MoSe$_2$ monolayer. (b) PL intensity *versus* sample position for a MoSe$_2$ monolayer. The position represents a path along the dashed line in panel (a).

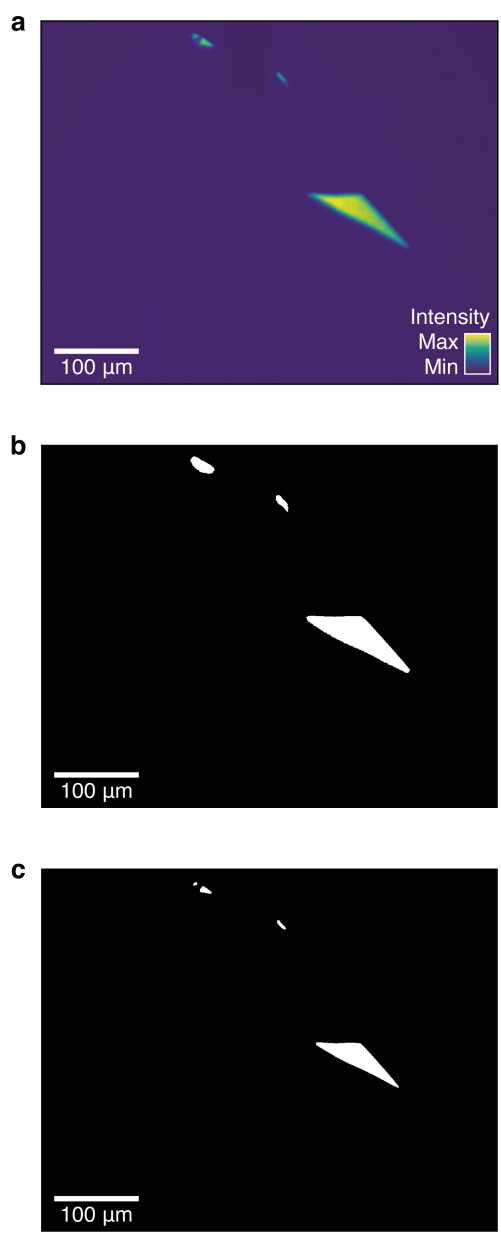

**Figure S9.** Flake analysis. (a) Photoluminescence image of exfoliated MoSe$_2$ on PDMS showing several monolayers. (b) Binarized image, obtained by applying a threshold to the photoluminescence image in (a). (c) Binarized image, computed by thresholding each individual monolayer of the image in panel (b). See also Methods.

# S2. SUPPLEMENTARY TABLES

| Abbreviation | Definition | Details |
|---|---|---|
| | White LED | Thorlabs, MCWHLP1 |
| | Camera (color) | Thorlabs, CS126CU |
| | Camera (monochromatic) | Thorlabs, DCC1545M |
| LP1 | Longpass filter 1 | 400 nm longpass filter |
| LP2 | Longpass filter 2 | 700 nm longpass filter, Thorlabs, FELH0700 |
| BP | Bandpass filter | 650 nm – 45 nm bandpass filter |
| SP | Shortpass filter | 650 nm shortpass filter, Brightline, 650/SP |
| | | 600 nm shortpass filter, Thorlabs, FES0600 |
| M1, M2, M3, M4 | Mirror | Thorlabs, BB1 – E02 |
| L1 | Lens 1 | Thorlabs, AC254-050-A-ML |
| L2 | Lens 2 | Thorlabs, AC254-300-A-ML |
| L3 | Lens 3 | Thorlabs, AC254-100-A-ML |
| L4 | Lens 4 | Thorlabs, AC254-100-A-ML |
| L5 | Lens 5 | Thorlabs, AC254-030-A-ML |
| L6 | Lens 6 | Thorlabs, AC254-100-A-ML |
| L7 | Lens 7 | Thorlabs, LA1509-A |
| L8 | Lens 8 | Thorlabs, AC254-50-A-ML |
| L9 | Lens 9 | Thorlabs, AC254-100-A-ML |
| BS1 | Beamsplitter 1 | Thorlabs, CCM1-BS013/M |
| BS2 | Beamsplitter 2 | Thorlabs, BSW10R |
| DM | Dichroic mirror | Thorlabs, DMLP505R |
| RM | Removable mirror | Thorlabs, PFR10-P01 |
| MO | Microscope objective | Nikon, 10×, NA 0.45, MRD00105 |
| MA1, MA2 | Motorized Actuator | Thorlabs, Z812B |
| C1, C2 | Controller | Thorlabs, KDC101 |

**Table S1.** Parts list for our setup, including an overview of the abbreviations and their definitions, as well as detailed information about the parts. The setup is depicted in Figure S3.

| From Tape | To Tape |
|---|---|
| 1 | 2 |
| 1 | 3 |
| 1 | 4 |
| 2 | 5 |
| 2 | 6 |
| 2 | 7 |
| 3 | 8 |
| 3 | 9 |
| 4 | 10 |
| 5 | 11 |
| 5 | 12 |
| 6 | 13 |
| 8 | 14 |
| 11 | 15 |

**Table S2.** Exfoliation scheme. Tapes are numbered from 1 to 15. According to the scheme, the "from" tape is brought in contact with the "to" tape. More information is provided in Methods.